\documentclass[preprint,12pt]{elsarticle}

\usepackage{amssymb}
\usepackage{amsmath}
\usepackage{amssymb}
\usepackage{yfonts}
\usepackage{amsmath}
\usepackage{verbatim}

\usepackage[utf8]{inputenc}
\usepackage{nicefrac}
\usepackage{svg}
\journal{Optik}

\begin{document}

\begin{frontmatter}



\title{Generalized Finite Differences Method Applied to Finite Photonic Crystal}


\affiliation[label1]{organization={Grupo de Física Atómica y Molecular, Instituto de Física, Universidad de Antioquia},
             city={Medellín},
             country={Colombia}}

\affiliation[label2]{organization={Grupo de Óptica y Fotónica, Instituto de Física, Universidad de Antioquia},
             city={Medellín},
             country={Colombia}}

\affiliation[label3]{organization={Grupo de Superconductividad y Nanotecnología, Departamento de Física, Universidad Nacional de Colombia},
             city={Bogotá},
             country={Colombia}}

\author[label1]{Santiago Bustamante\corref{c1}}
\cortext[c1]{Corresponding author}
\ead{santiago.bustamanteq@udea.edu.co}
\author[label2]{Esteban Marulanda}
\author[label1]{Jorge Mahecha}
\author[label3]{Herbert Vinck-Posada}

\begin{abstract}
We propose a Generalized Finite-Differences in the Frequency Domain method for the computation of photonic band structures of finite photonic crystals. Our approach is to discretize some fundamental domain instead of a single unit cell, such that boundary conditions of interest can be introduced into the eigenvalue problem. The validity and effectiveness of the proposed method are shown for the case of a one-dimensional photonic crystal embedded in an optical cavity. The limit from finite to infinite photonic crystals is reviewed in view of the proposed method.

\end{abstract}


\iffalse
\begin{highlights}
\item We propose and demonstrate a novel generalized method for the computation of photonic band structures.
\item We compute the photonic band structure of a one-dimensional finite photonic crystal inside an optical cavity using the proposed method.
\item We review the limit from finite to infinite photonic crystals in view of the proposed method.
\end{highlights}
\fi


\end{frontmatter}


\numberwithin{equation}{section}
\section{Introduction}

In recent years, photonic crystals (PCs) have been demonstrated to exhibit unique properties that allow control of the light flow. These materials, characterized by their optical periodicity, have found widespread applications in various fields. For instance, PCs are used in realizing quantum optical phenomena \cite{PhysRevA.50.1764, PhysRevB.43.12772}, designing quantum circuits \cite{OBrien2009}, supercontinuum generation \cite{2010}, fabrication of solar cells \cite{Liu2019}, biomedical and biological applications \cite{Chen2017,Lv2023}, etc. On the other hand, the study and scope of applications of PCs are largely linked to proposing numerical solution methods. In literature, there are a vast number of methods for calculating PC properties, some of the newest being based on machine learning \cite{Ma2020,Muhammad2022}. Among the most widely known methods include the plane wave method (PWE) \cite{Sakoda2005}, the finite difference method in the time domain (FDTD) \cite{Taflove2005-fh}, and the finite difference method in the frequency domain (FDFD) \cite{Ohtaka2004}. 

% Comprobar un poco las referencias, ¿por qué colocariamos esas aplicaciones?. Cuando se introducen los metodos numericos, ¿por qué estos? almenos se deberia de recalcar que son populares dentro del contexto de analisis.

%The implementation of most of these methods rely on the spatial periodicity of the PCs, which is assumed to be infinitely extended throughout space. This property makes it possible to solve the computational problem in a single-unit cell, considering Bloch's theorem. Even when the system possesses a defect and does not exhibit strict periodicity, its properties can be approximated by the supercell method \cite{PhysRevE.67.026612}. The efficiency of the method is directly related to the size of the supercell.

The implementation of most of these methods relies on the validity on Bloch's theorem, which is expected from the spatial periodicity and infinite spatial extension of the crystal. This theorem makes it possible to solve the computational problem in a single unit cell. Even when the system possesses a defect and does not exhibit strict periodicity, its properties can be approximated by the supercell method \cite{PhysRevE.67.026612}, whose efficiency is directly related to the size of the supercell.

%Those methods usually have aSolution for PC becomes relatively easy since require the specification of unit cell and application of Bloch condition, which assumes periodic crystals and extends through the whole space.
%The PWE method usually has problems for large systems since the algorithm's complexity is $O(N^3)$. Moreover, the FDTD method depends on the number of iterations in time \cite{Guo:04}, which demands much computational cost. The FDFD method, on the other hand, has advantages such as fast convergence of the solution and stability. 

% Cada método tiene sus ventajas y desvenjas. Este parrafo intenta motivar porque en el articulo se esta utilizando el método FDFD como método de analisis y no otro. Por lo que debe de quedar muy claro y en detalle esto ultimo. Debería de almenos buscar 3 referencias que comprueben el punto.

However, recent technological advances that involve reducing systems to a nanoscopic scale or modeling realistic devices \cite{2014NaPho...8..406H,Inoue2022} make Bloch's theorem less rigorously applicable. For this reason, different methods have been studied to approach the finite PC problem. These methods are derived from rigorous treatments of the finite problem \cite{Pereyra2021} or the use of techniques for infinite systems applied to the finite case \cite{Chen:22} (usually considering Bloch waves). 

% Este parrafo introduce el problema principal del articulo. Esto es, ¿como utilizar herramientas de solución de sistemas infinitos para sistemas finitos? Para ello es bueno motivar porque utilizar 

Given the relevance of Bloch's theorem and the advantages provided the FDFD method \cite{Guo:04}, it becomes valuable to identify the circumstances under which Bloch-type solutions remain valid in finite PCs since, in this case, they are not exactly eigenfunctions of the problem \cite{Ivchenko2012-xv}. In this paper, we propose a novel method for the photonic band structure computation of finite PCs based on the validity of Bloch's theorem in finite systems. The proposed method expands the scope of the FDFD method by allowing the introduction of more general boundary conditions. This enables photonic band gap analysis in the limit from finite to infinite PCs. We illustrate the effectiveness and validity of this method by examining the case of a one-dimensional finite PC inside an optical cavity.

\section{Generalized FDFD Method} \label{sec:gfdfd_method}
\begin{figure}[h]
    \centering
    \includegraphics[width=8 cm]{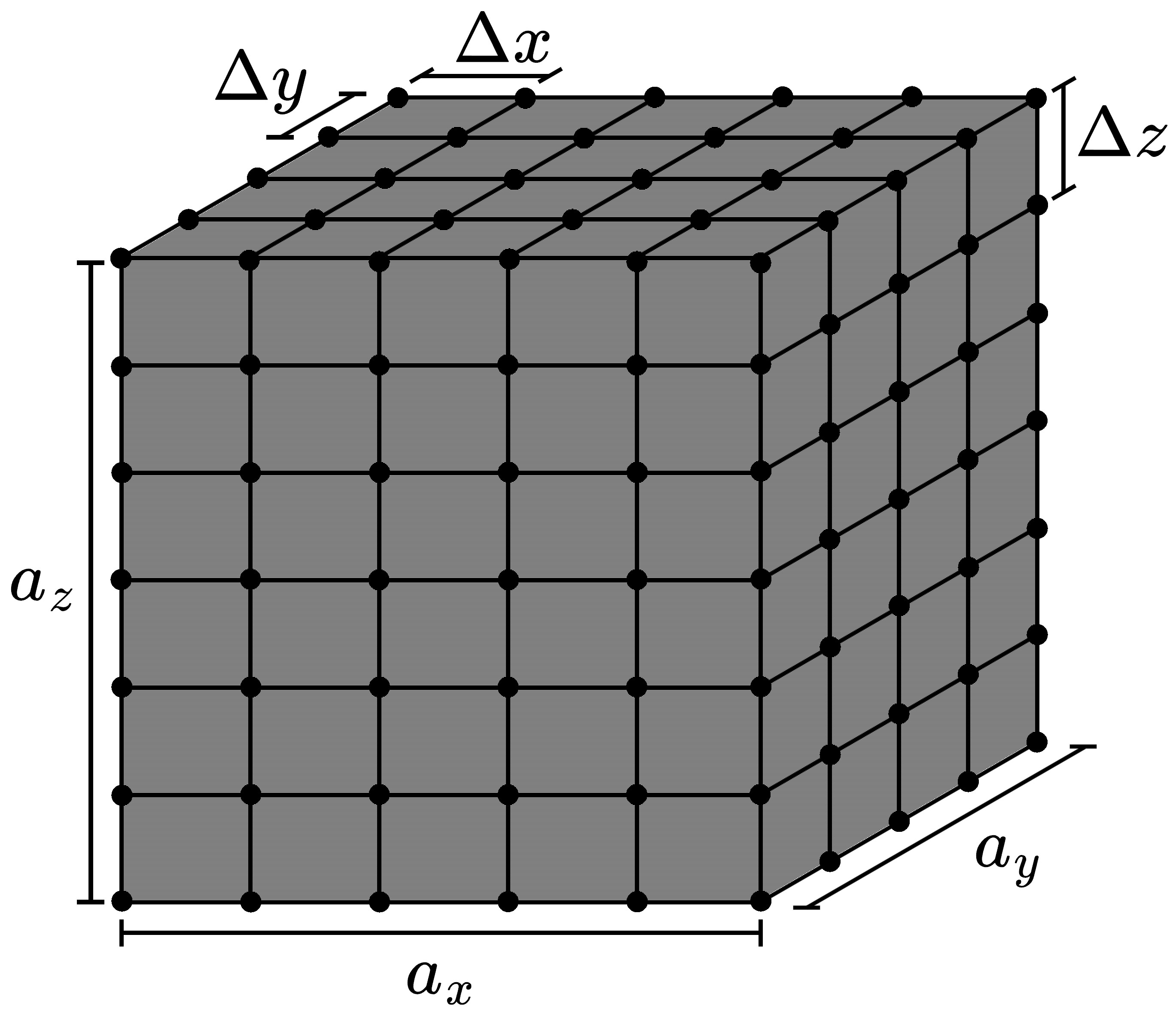}
    \caption{Sampling grid in a rectangular unit cell. Here $a_x$, $a_y$, and $a_z$ are the lattice parameters of the crystal and $\Delta x$, $\Delta y$ and $\Delta z$ are the sampling resolutions in each spatial axis. This discretization of space allows one to approximate linear differential operators by finite matrices up to a truncation error, which decreases with smaller sampling resolutions.}
    \label{fig:unit_cell}
\end{figure}

\indent Before introducing the Generalized FDFD (GFDFD) method, let us briefly recall some things about the FDFD method. The FDFD method is a finite-differences method for the computation of PC eigenmodes that even allows supercell techniques to study crystal defects in the lattice \cite{Vasco2010}. This method reduces the problem of solving Maxwell equations in periodic dielectric media to the diagonalization of a single finite matrix.
To quickly summarize how the method works, consider the decoupling of the time-independent Maxwell's equations for the electric and magnetic fields
\begin{subequations}\label{eq:eigenvalue_problem}
\begin{align}
        \mathcal{L}_E \mathbf{E} &\triangleq \frac{1}{\epsilon}  \nabla \times \left( \nabla \times \mathbf{E} \right) =  \frac{\omega^{2}}{c^{2}} \mathbf{E},\label{eq:maxwell1}\\
        \mathcal{L}_H \mathbf{H} &\triangleq \nabla \times \left(\frac{1}{\epsilon}  \nabla \times \mathbf{H} \right) = \frac{\omega^{2}}{c^{2}} \mathbf{H},
        \label{eq:maxwell2}
\end{align}
\end{subequations}
\noindent where $\mathcal{L}_E$ and $\mathcal{L}_H$ are linear differential operators on the fields, $\epsilon$ is the relative dielectric function characterizing the PC of interest, $\omega$ is the angular frequency of the fields and $c$ is the speed of light. These equations define a single eigenvalue problem from which one can obtain a set of eigenmodes, each with its corresponding eigenfrequency and field spatial profiles. Choosing to solve the eigenvalue problem for either $\mathcal{L}_E$ or $\mathcal{L}_H$ is a matter of convenience. However, the operator $\mathcal{L}_H$ is usually chosen since it is Hermitian, and the transversality condition for the magnetic field $\nabla \cdot \mathbf{H} = 0$ is easier to work with \cite{Joannopoulos:08:Book}. Once an operator $\mathcal{L}$ is chosen, we define a sampling grid with $n$ points within a single unit cell of the crystal that allows one to represent the differential operator $\mathcal{L}$ by a $n\times n$ matrix $L$ up to an error due to discretization of space (see Figure \ref{fig:unit_cell}). In principle, this truncation error can be arbitrarily reduced by using finer grids.

To properly define $L$, however, boundary conditions on the unit cell are needed. These are supplied by the fact that solutions to \ref{eq:eigenvalue_problem} can be labeled by a wave vector $\mathbf{k}$ in the First Brillouin Zone (FBZ) and satisfy the equations
\begin{subequations} \label{eq:blochwavecondition}
\begin{align}
    \mathbf{E}_\mathbf{k} (\mathbf{r} + \mathbf{a}) & = e^{i \mathbf{k} \cdot \mathbf{a} }  \mathbf{E}_\mathbf{k} (\mathbf{r}) \\
    \mathbf{H}_\mathbf{k} (\mathbf{r} + \mathbf{a}) & = e^{i \mathbf{k} \cdot \mathbf{a} }  \mathbf{H}_\mathbf{k} (\mathbf{r})
\end{align}
\end{subequations}
\noindent for any arbitrary crystal lattice vector $\mathbf{a}$. The wave vector $\mathbf{k}$ is generally known as the crystal momentum. Equations \ref{eq:blochwavecondition} are also known as Bloch's theorem or Bloch's wave condition and are only valid if $\epsilon$ is a periodic function that extends through all space. Since this is the conventional way PCs are defined, the Bloch's theorem can be naturally used and does not represent an actual constraint.
Once $L$ is adequately defined, its eigenvalues and eigenvectors can be computed by modern numerical diagonalization methods, thus yielding an excellent approximation to the photonic band structure of the crystal and its corresponding eigenfunctions.

The FDFD method, as described above, allows one to study the optical properties of PCs that are infinitely extended through all space. Although, in reality, this is a physically impossible situation, methods based on this theoretical assumption still provide a realistic description of PCs. This is due to the fact that PCs are built such that the size of their unit cells is considerably small compared to the size of the crystals themselves. However, one can still ask for the possibility of studying periodic dielectric structures that do not necessarily extend endlessly through infinite space, such as real PCs embedded in optical cavities. With that in mind, we now propose a generalization of the FDFD method based on a slightly different discretization of space that allows a more general manipulation of the boundary conditions of the eigenvalue problem defined in Equations \ref{eq:eigenvalue_problem} while forcing Bloch's wave condition. We will see how this generalization allows one to study Bloch waves in finite systems.

Our proposed generalization of the FDFD method goes as follows. We start by specifying a sampling grid with $N$ points within some fundamental domain instead of a single unit cell. This domain must comprise a finite amount $M$ of complete non-overlapping unit cells, each with $n$ sampling points, and is defined to include the boundary conditions of interest. For instance, to find the eigenmodes of the radiation field in a PC embedded in a resonant cavity, the whole cavity must be spatially sampled to apply reflective boundary conditions to the field. Again, this discretization of space and the boundary conditions allow us to represent the linear differential operator $\mathcal{L}$ as an $N\times N$ matrix $L$.

\begin{figure}
    \centering
    \includegraphics[width=13 cm]{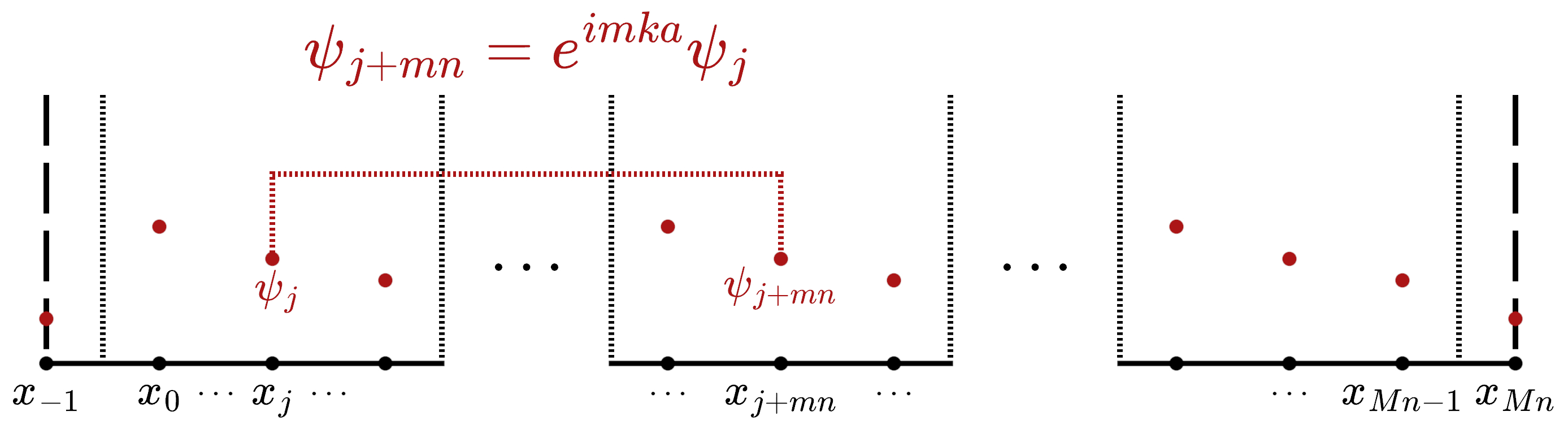}
    \caption{One-dimensional representation of a fundamental domain with $M$ unit cells and lattice constant $a$. Dashed and dotted dark vertical lines indicate the boundaries of the fundamental domain and the unit cells respectively. The field evaluated at point $x_j$ is denoted by $\psi_j$. Boundary conditions are applied at $x_{-1}$ and $x_{Mn}$. A red dashed line indicates the Bloch's wave condition ($j=0,1,\dots,n-1$ and $m=0,1,\dots,M-1$), such that only $n$ values of the field are actually independent.}
    \label{fig:bloch_enforcement}
\end{figure}

At this point, the eigenvalue problem can already be solved numerically. However, this may represent a great computational cost if either $M$ or $n$ are too large, and the solutions will not be, in general, Bloch waves. In order to filter out non-Bloch-type solutions and reduce the computational cost, we enforce the Bloch's wave condition by expressing the field in all unit cells as phase-shifted copies of the field in a single unit cell (see Figure \ref{fig:bloch_enforcement}). This constraint reduces the number of independent variables from $N$ to $n$ and consequently the matrix $L$ can be transformed into an $n\times n$ matrix $\tilde L = B^{-1} L B$, where $B$ is defined as a $N \times n$ matrix by the equation
\begin{equation}\label{eq:bloch_matrix}
    \left[
\begin{array}{c}
\psi_0 \\
\psi_1 \\
\vdots \\
\psi_{N-1} \\
\end{array}\right] = \sqrt{M} B \left[
\begin{array}{c}
\psi_0 \\
\psi_1 \\
\vdots \\
\psi_{n-1} \\
\end{array}\right],
\end{equation}
and $B^{-1}$ is the left pseudo-inverse of $B$, which is simply its conjugate transpose. In principle, the boundary and the Bloch's wave conditions are included in $\Tilde{L}$, so the diagonalization of this matrix will yield the eigenmodes of the embedded PC. That is the essence of the GFDFD method. When $M=1$ and periodic boundary conditions are used, we return to the FDFD method. In this sense, the FDFD method is a particular case of the GFDFD method.

\section{Photonic crystal in an optical cavity} \label{sec:method_compar.}
% en ese caso la solucion podria ser periodica en ese caso tal como GFDFD.

To explicitly illustrate the method, the case of a one-dimensional finite bilayer crystal embedded in a cavity with metallic boundary conditions is studied. We considered a transverse mode of the electric field, in which case the discretization of Equation \ref{eq:maxwell1} takes the form

\begin{equation} \label{eq:1dfdfd}
\sum_{j=0}^{n-1} L_{ij} E(x_j) = - \frac{\omega^2}{c^2} E(x_i),
\end{equation}

\noindent where $ L_{ij} = (\delta_{i+1,j}-2\delta_{ij} + \delta_{i-1,j})/(\epsilon (x_j)\Delta x ^2)$ and metallic boundary conditions are imposed by setting $\delta_{-1,j},\delta_{Mn,j} \rightarrow 0$. The diagonalization of the $Mn\times Mn$ matrix $L$ provides a complete set of eigenmodes of the radiation field in the cavity. These modes are computed to be later compared to those provided by the GFDFD method, which is implemented by the diagonalization of the $n\times n$ matrix  $\tilde L = B^{-1} L B$, where

\begin{equation}
     B=\frac{1}{\sqrt M}\left[
 \begin{array}{c}
 I_n \\
 e^{ika}I_n \\
 \vdots \\
 e^{i(M-1)ka}I_n \\
\end{array}\right],
\end{equation}

\noindent $I_n$ being the $n\times n$ identity matrix. At this stage, it is important to note that the GFDFD method relies on the validity of Bloch's theorem in the fundamental domain. Consequently, all solutions provided by this method will exhibit periodicity in their amplitudes across all unit cells. This means only a subset of the complete set of solutions can be obtained for finite systems using the GFDFD method. This subset is specified by the node theorem as follows: let $\omega_r$ represent the $r$-th eigenvalue of the Sturm-Liouville problem defined by Equation \ref{eq:1dfdfd}. If the eigenvalues are labeled in increasing order ($\omega_1<\omega_2<\dots$), then it is well established that the $r$-th eigenfunction $\psi_r$ has exactly $r-1$ nodes \cite{teschlordinary}. Therefore, for a finite PC with $M$ unit cells, the eigenmode $\psi_{r}$ of $L$ can only exhibit periodic amplitude among all cells if $r$ is a multiple of $M$. This lets us filter out most non-Bloch-type solutions, allowing a more straightforward comparison between the eigenmodes of $L$ and $\tilde L$.

\begin{figure}[ht]
\centering
\includegraphics[width=1\linewidth]{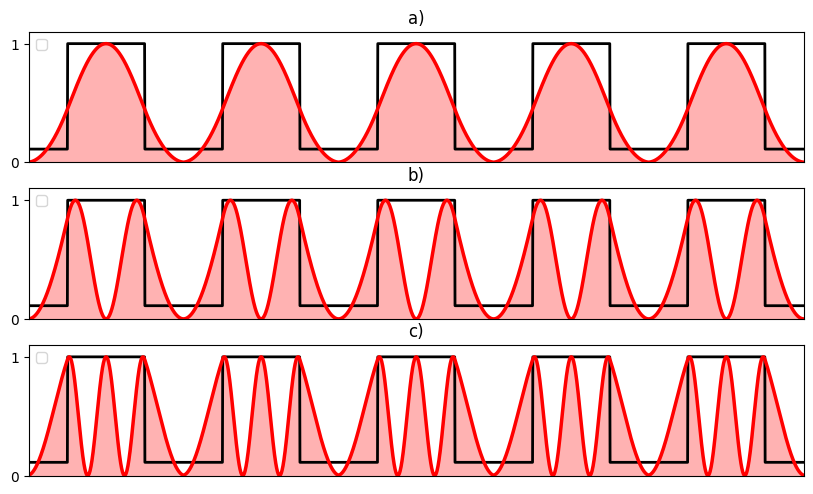}
\vspace{0.5cm}
\caption{Red lines represent the normalized intensity profiles of the obtained solutions. (a), (b) and (c) correspond to the fifth, tenth, and fifteenth eigenmodes, respectively. The black line indicates the normalized dielectric profile.}
\label{c1}
\end{figure}

\begin{figure}[ht]
\centering\includegraphics[width=14 cm]{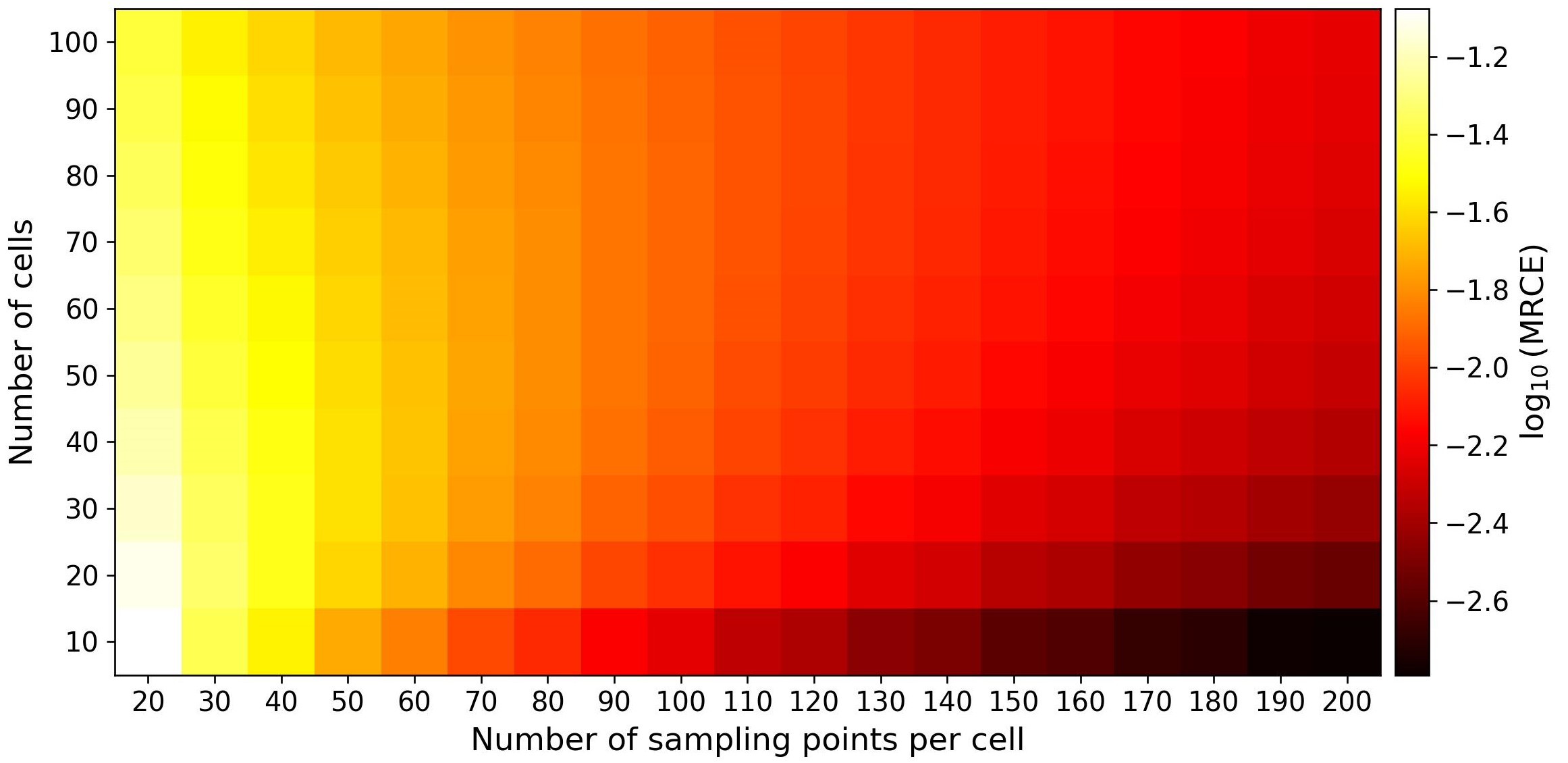}
\vspace{0.5cm}
\caption{Convergence colormap of the GFDFD method. Color represents the logarithm of the MRCE. The method appears to converge for all number of cells.}
\label{convergence}
\end{figure}

The intensity profiles are presented in Figure \ref{c1} for the case of $M=5$ unitary cells. For this computation, we used $n=512$ sampling points per unit cell, a lattice constant of $a=1$, a wave number of $k=0$, and relative dielectric constants of 1 and 9. The profiles obtained from $\Tilde{L}$ were almost indistinguishable to those given by $L$. This means that the GFDFD provides meaningful information about the actual eigenfrequencies of the system, even when Bloch's theorem does not necessarily hold. Similar results are found for a higher number of unit cells, and the convergence of the method seems to depend mostly on the number $n$ of sampling points per unit cell. In Figure \ref{convergence} we show how the mean relative convergence error (MRCE) of the first mode eigenfrequency, defined by
\begin{equation} \label{eq:MRCE}
\text{MRCE} = \frac{a_0}{2\pi} \int_{\text{FBZ}} dk \left | \frac{\omega^{(n)}(k) -\omega^{(n-10)}(k)}{\omega^{(n)}(k)} \right |,
\end{equation}
approaches zero as $n$ increases for all $M$, where $\omega^{(n)}(k)$ is the eigenfrequency of the first mode computed with $n$ sampling points per unit cell. The method converges quicker for lower values of $M$ and similar convergence results are obtained for higher modes.

\begin{figure}[ht]
\centering\includegraphics[width=13 cm]{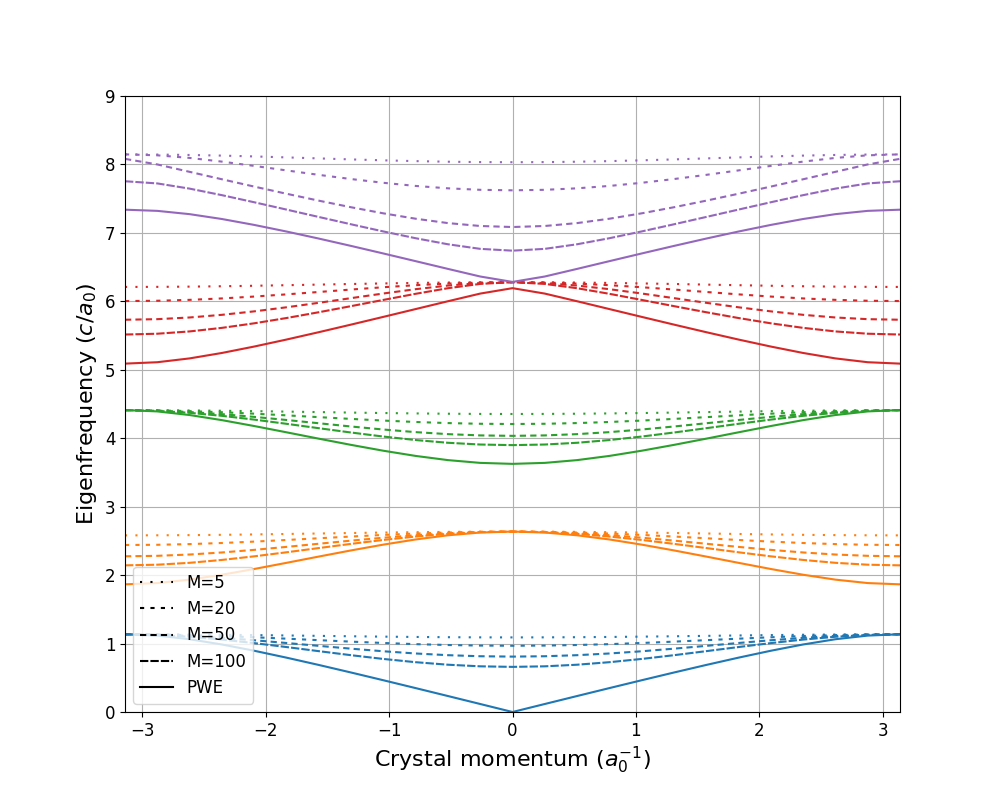}
\vspace{0.5cm}
\caption{Photonic band structures of the one-dimensional infinite and finite bilayer PCs. Solid and dashed lines represent the solutions provided by the PWE and GFDFD methods for the infinite and finite systems respectively. The band structures were computed with 180 sampling points per unit cell.}
\label{c2}
\end{figure}

To further study the validity of the method and the optical properties of PCs in the finite-to-infinite limit, we computed the photonic band structures of the system for increasing $M$ values. The band structure for a 1D infinite bilayer PC with the same lattice and dielectric constants is also computed using the PWE method. The results behave as expected: as $M$ increases, the band structure of the finite system converges to the one of the infinite system (see Figure \ref{c2}). This is because as the number of cells increases, the ratio between the length of the crystal and the lattice constant increases as well, making the crystal effectively infinite compared to the dimensions of its unit cells. This convergence suggests that the crystal momentum, which is initially introduced by the enforcement of Bloch's wave condition, acquires more physical relevance as the size of the crystal increases.

% CRISTAL DE DOS CAPAZ.

\section{Conclusions} 

We proposed and demonstrated a GFDFD method for the computation of the photonic band structures of finite PCs. This method, which strongly relies on the validity of Bloch's theorem in finite systems, allows the calculation of Bloch-type solutions in finite systems. We particularly showed the method's validity and efficacy by studying the eigenmodes of a one-dimensional finite bilayer PC embedded in an optical cavity. The intensity profiles of the radiation fields given by the GFDFD method showed a high resemblance to those obtained before using Bloch's theorem, thus providing meaningful information about the actual eigenfrequencies of the composite system, with the addition of a new parameter of the set of solutions, namely the crystal momentum. This parameter, introduced by the enforcement of Bloch's wave condition, appears to acquire more physical relevance as the size of the crystal increases, as suggested by the photonic band structure analysis in the finite-to-infinite limit. In this limit, the photonic band structure given by the GFDFD method for the finite PC appears to converge to that provided by the PWE method for the infinite PC, just as expected. This further solidifies the validity of the proposed method. As a perspective, we want to provide experimental support to the GFDFD method. We are also looking forward to explicitly demonstrate the validity of the method in two dimensions, showing particular cases of finite PCs in manifolds with non-trivial topologies, such as the Möbius Strip or the Klein Bottle, where boundary conditions on the fundamental domain are determined by the corresponding quotient space \cite{Bělín_2021}.

\bibliographystyle{unsrt}
\bibliography{ref}

\end{document}